Wiktor Bohdan DASZCZUK[1], Jerzy MIEŚCICKI[2]

# Distributed management of Personal Rapid Transit (PRT) vehicles under unusual transport conditions

**Abstract.** The paper presents a flexibility of management of vehicles in Personal Rapid Transit (PRT) network. The algorithm used for delivering empty vehicles for waiting passengers is based on multiparameter analysis. Due to its distributed construction, the algorithm has a *horizon* parameter, which specifies the maximum distance between stations the communications is performed. Every decision is made basing on an information about situation (number of vehicles standing at a station, number of vehicles travelling to a station, number of passengers waiting) sent between stations, without any central data base containing traffic conditions. The simulation of the traffic in random case (typical) and in unusual case of delivering people to a social event occurring at single place is presented. It is shown that simple manipulation with *horizon* parameter allows to adapt the network to extremely uneven demand and destination choice.

**Keywords:** Personal Rapid Transit; Urban Transport; Transport Simulation; Transportation Management

## INTRODUCTION

Personal Rapid Transit (PRT) is innovative urban transit system [1,2,7,8] organized as a network covering, for instance, a part of the city, fairgrounds, multi-terminal airport, etc. The driverless (i.e. fully system-controlled) vehicles move along one-way tracks, separated from a conventional traffic e.g. due to elevation of tracks above the ground level. Typically, vehicle can carry a group of 1 up to 6 passengers. Term 'personal' emphasizes that a passenger (or group of passengers) can choose freely the time as well as the destination of a trip instead of being forced to follow a strict, predefined timetable.

The system determines the best route for the trip (which is not necessarily the shortest one) and controls the vehicle movement during the voyage (acceleration/deceleration, preserving the separation between vehicles, passing the junctions, avoiding traffic jams etc.). System is also responsible for the management of empty vehicles: sending them where (and when) they are needed, expelling them from a station whenever the free place is needed for new, arriving ones etc.

The original multi-parameter algorithm for PRT traffic management was elaborated [5] within the ECO-Mobility project [6], done in Warsaw University of Technology and co-financed from European Regional Development Fund/Innovative Economy Operational Programme. The algorithm has proved useful and effective in many simulations performed in Feniks simulation environment [4], developed under the same project. Many variants of the algorithm have been tested against several models (differing in topography, the intensity and profile of demand, fleet size and other factors) in various traffic conditions [4,5].

---

[1] Warsaw University of Technology, Institute of Computer Science, Nowowiejska Str. 15/19, 00-665 Warsaw, Poland, wbd@ii.pw.edu.pl
[2] Warsaw University of Technology, Institute of Computer Science, Nowowiejska Str. 15/19, 00-665 Warsaw, Poland, jms@ii.pw.edu.pl

Generally, the PRT management algorithm is used for:
– routing vehicles to their destinations,
– dynamic routing (adapting the route to changing traffic conditions, e.g. avoiding the traffic jams),
– calling vehicles for new passengers (if at a given station there are no empty ones at the moment),
– expelling empty vehicles (resting at the station) to make room for approaching full vehicles (if at a given station there are currently no free berths to disembark the passengers),
– balancing the distribution of empty vehicles over the network in order to make them more easily accessible for the future use,
– withdrawing empty vehicles to capacitors (e.g. for safety reasons, cleaning, maintenance purposes etc.).

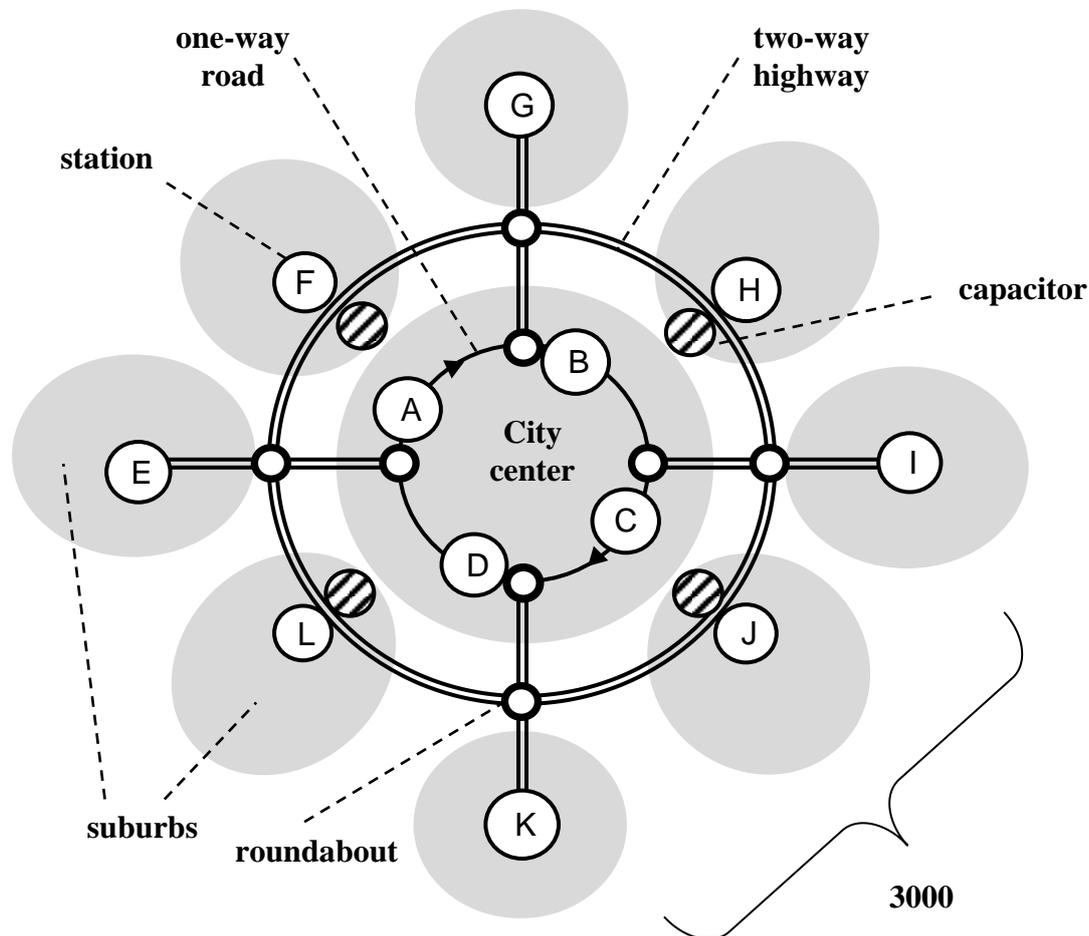

**Fig. 1.** The *City* model

The management algorithm is built as a target function with several coefficients and thresholds, depending on: numbers of empty and occupied berths at the stations, distance and traffic conditions between individual stations, passenger queue lengths and numbers of full/empty vehicles travelling to individual stations.

One advantage of the prepared algorithm consists in its distributed construction. No centralized representation of global network state is needed and all decisions are based on local exchange of information between stations and between stations and vehicles. Maximum distance in which such a local communication takes place – is limited by the *horizon*, which is one of parameters of the management algorithm. Its meaning is the following.

Distance is, of course, the shortest path between two stations. For a given network we can build whole matrix of distances between pairs of stations. This matrix is not symmetric because tracks are unidirectional. Having such a matrix we can compute an average inter-station distance (AISD). This is used as a normalizing constant: the horizon is the ratio of an absolute distance and AISD. So, for instance, setting horizon equal to 0.5 causes that only the data from network stations located no more than one half of AISD are taken into account, etc. Similarly, no horizon declared means that any station communicates with all other stations. This feature frees the designer from expressing horizon in absolute units.

Numerous simulation experiments have confirmed the effectiveness of distributed vehicle management algorithm. However, the research has been aimed mainly to PRT network behavior under the balanced load. In particular, the number of vehicles was determined so that (for a given daily or hourly demand profile) the equilibrium state of network was practically guaranteed, even if due to increased passenger arrival rate (e.g. during rush hours) the waiting queues as well as the waiting time may temporarily grow.

On the other hand, any transport system has to deal with rare but challenging situations when the demand suddenly and significantly exceeds the margins of assumed network performance. Fire or other emergency, attractive sport events, rock concerts, public rallies etc. – are good examples of increased demand to travel to (and from) some particular place. Usually, the operation of transportation system has to be supported by additional human intervention: changes of traffic organization, additional bus lines etc. But to what extent the system-controlled PRT is capable to adapt itself to such a challenging situation?

Below, we show that the horizon is a convenient parameter that can be useful in tuning the management algorithm. By changing the decision horizon for a time, we can make the network to largely accommodate the unusual traffic conditions.

## 1. TRANSPORT TASK

To support the intuition let us assume that a social event is prepared in a single location in the city (a stadium for example). The participants should be moved from many places to the event area, and after the event the participants should be ably moved back home. By assumption, the participants begin travelling to the event area two hours before its beginning, and they leave the event area gradually during two hours after its end.

For the experiment, a *City* model has been chosen (Figure 1). It is one of proposed PRT benchmarks described in [9]. The event takes place near the *I* station on the right edge of the model. The total population of *City* is 33750 inhabitants, which is computed from the assumed size of dashed areas and the typical population density of an average Polish city [10]. For the purpose of the experiment assume that every eighth person is going to take part in the event (>4000 participants).

The model has 12 stations, which for 4219 people gives total 191.8 persons/h entering every of 11 stations except the target station. Also, it is assumed that there exists normal, "background" traffic between stations (not associated with the event), but limited to 4 transit orders/h at every station (including the event area station). These trips are executed to randomly chosen target stations. To sum up, two hours before the event, passengers start to arrive at 11 stations every 0.856 minute (on the average), and at the event area station every 15 minutes.

In both cases, the inter-arrival time is the random variable. Its distribution is exponential with the mean value equal to 0.856 min and 15 min, respectively. In other words, the arrival is the Poisson process with rate $\lambda$ which is a reciprocal of the mean inter-arrival time.

In the simulator, default cardinality of passenger groups travelling in a single vehicle has uniform distribution: 1-, 2-, 3- or 4-passenger groups, each with the probability equal to 0.25. We have assumed that on a way to the event the passengers will travel in larger groups. The distribution is no longer uniform but the following: 1 person-10%, 2-20%, 3-40%, 4-30%, which gives average cardinality of 2.9 people/trip. Therefore, any of 11 stations receives 66.13 groups/h.

It is assumed that the participants leave the event area (when it finishes) gradually during two hours. As the participants are all at one station, it is easier to complete larger groups: 1-5%, 2-5%, 3-30%, 4-60%, which gives average group cardinality 3.45 persons/trip. As a result, 611.4 groups of participants per hour arrive at station *I*, and taking into account 4 groups/h of background traffic, a group orders the transit at the station *I* every 0.0976 minute (average, with exponential distribution).

## 2. SIMULATION RESULTS

### 2.1. Uniform demand and random destination

The model was tested first with exponential demand distribution, average value 0.856 (as described above), but with uniform origin-destination matrix (travel to every other station is equally probable).

The purpose was twofold: firstly, to make the base for comparison, second - to check if the network ridership is enough for such demand. Various horizon values were tested, from no horizon (a station communicates with all other stations), to 0.5 (half the AISD). The output parameters are – AWT (average waiting time – the time of passenger groups waiting at a station for a vehicle to be delivered), AQL – average queue length, maxQL – maximum queue length in the experiment. The results are collected in Table 1.

The results show that all three values of the horizon give comparable results, but all of them are better than in "no horizon" case. It is important, because the higher the horizon value the greater volume and cost of the communication. Horizon = 0.5 is the best as it gives reasonable results at rather small communication cost.

**Tab. 1.** Main output parameters, random case

| Horizon | AWT [s] | AQL [groups] | maxQL [groups] |
|---|---|---|---|
| no horizon | 73.2 | 1.44 | 11 |
| 1.5 | 60.9 | 1.06 | 7 |
| 1.0 | 61.5 | 1.16 | 12 |
| 0.5 | 61.5 | 1.16 | 12 |

### 2.2. Delivering participants to the event area

The second experiment has been performed with bringing participants to the event area. Origin-destination matrix reflects the traffic directed to the event area and some trips (4/h) directed to randomly chosen destinations. The results are collected in Table 2. The new, additional column "Rest" shows how long it takes to deliver passengers waiting in queues while the demand has stopped (demand lowered to 4 groups/h at all stations).

**Tab. 2.** Main output parameters, delivering participants

| Horizon | AWT [s] | AQL [groups] | maxQL [groups] | Rest [min] |
|---|---|---|---|---|
| no horizon | 719.0 | 7.6 | 73 | 31 |
| 1.5 | 548.0 | 5.9 | 66 | 28 |
| 1.0 | 670.4 | 6.5 | 96 | 35 |
| 0.5 | 2096.9 | 44.7 | 131 | >2h |

Horizon = 0.5 is not acceptable in this case, as maxQL rises to 131 groups, AWT is greater than half an hour, and it is not possible to deliver the participants during 2 hours after the demand drops down. Enlarging the horizon to 1 is enough to acquire reasonable network work: 11 minutes of waiting time at a station and delivering all participants in about half an hour after the demand drops. The horizon =1.5 gives even better results, but at the cost of higher communication overhead.

The Rest ≈ 0.5h shows that the organizers should inform the participants that it is better to start from home not 2 hours, but rather 2.5 hours before the event starts.

## 2.3. Bringing participants back home

The results for bringing passengers back home are collected in Table 3. This time, AWT and AOL are shown for the station next to the event area.

For horizon 0.5 the network is almost blocked: passengers wait at station *I*, but the network delivers empty vehicles so slowly that the passengers are not shipped even through 2 hours after the demand drops. The difference between horizons 1 and 1.5 is large (more than twice in AWT), therefore horizon 1.5 is most reasonable, even at the cost of communication growth.

**Tab. 3.** Main output parameters, shipping participants back home

| Horizon | AWT [s] | AQL [groups] | maxQL [groups] | Rest [min] |
|---|---|---|---|---|
| no horizon | 1010.3 | 58.5 | 247 | 35 |
| 1.5 | 1277.1 | 74.8 | 311 | 37 |
| 1.0 | 3038.7 | 186.3 | 561 | 97 |
| 0.5 | 5080.7 | 595.2 | 1197 | >2h |

## 3. CONCLUSIONS

A social event that attracts a significant part of the population always is a challenge for a transport system. Typical solution is to use additional means (e.g. special bus lines etc.) to perform the transport task.

In this paper a flexibility of PRT network is shown. Due to the algorithmic control, the problem can be solved (or at least mitigated) by simply changing the value of horizon – one parameter in distributed algorithm for vehicle management. The value 0.5, used in normal operation mode (with uniform demand and random destination choice), may be simply enlarged to value 1 while delivering the participants to the event area or to value 1.5 while shipping them back home. An interesting idea is that the decision on the horizon may be even performed automatically, based on the observation of queue lengths (which is equal to a number of unhandled transit orders).

The simulated model of a City is a theoretical construct, designed primarily as a benchmark for the research on management algorithms and simulation methodology. Also, the traffic data and parameters of the algorithm have been determined arbitrarily. Nevertheless, the research raises some valuable tips that can be useful when it comes to practical, real-life examples.